\documentclass{timproc}

\usepackage[utf8]{inputenc}
\usepackage{amsmath}
\usepackage{amsfonts}
\usepackage{amssymb}
\usepackage{graphicx}

\usepackage[figuresright]{rotating}
\usepackage{multirow}

\usepackage{hyperref}

\title[Circular periodic orbits and inclination excitation]{Circular periodic orbits, resonance capture and inclination excitation during type II migration}

\author[Antoniadou \& Voyatzis]{K. I. Antoniadou$^{1,2}$ and G. Voyatzis$^1$}
\affiliation{$^1$Aristotle University of Thessaloniki, Department of Physics, University Campus,\\ 54124 Thessaloniki, Greece\\
	\email{kyant@auth.gr}\\[1ex]
	$^2$University of Namur, Department of Mathematics, NaXys, 8 Rempart de la Vierge,\\ 5000 Namur, Belgium}

\begin{document}
\setcounter{page}{1}
\maketitle

\begin{abstract}
We consider planetary systems evolving under the effect of a Stokes-type dissipative force mimicking the outcome of a type II migration process. As inward migration proceeds and the planets follow the circular family (they start on circular orbits) and even though they are initially almost coplanar, resonance capture can be realized. Then, at the \textit{vertical critical orbits} (VCOs), that the circular family possesses, the  inclination excitation can abruptly take place. The planets are now guided by the spatial elliptic families, which bifurcate from those critical orbits. We herein, perform a direct link of mutually inclined stable planetary systems on circular orbits trapped in \textit{mean-motion resonance} (MMR) with the existence of VCOs of high values of multiplicity. It is shown that the more the multiplicity of the periodic orbits of the circular family increases, the more VCOs (corresponding to more MMRs) appear. In this way, we can provide a justification for the existence of resonant planets on circular orbits, which could, even further to that, evolve stably if they were mutually inclined.

\end{abstract}

\section{Introduction}
Among the exoplanets discovered so far, there are plenty of them located at small orbital distances ($a<0.2$ A.U.), where due to not only high temperature, but, also, insufficient amount of protostellar matter, accretion \textit{in situ} would have been prevented (\citealt{boss95,bode2000}). Apart from mechanisms, such as\vspace{-0.5em}
\begin{enumerate}
	\item gravitational interactions between two or more planets of Jovian masses, which possibly lead to the development of dynamical instability; This, in turn, could invoke orbit crossing and scattering of a planet leaving the other one in highly eccentric orbit and in some cases, this orbit has, also, small periastron distance that can circularize it through tidal dissipation (\citealt{rasio96})\vspace{-0.5em}
	\item effects of a distant companion star; In binary stellar systems dynamical instabilities and, therefore, chaos lead to highly eccentric orbits, which could, sometimes, induce collision between the planets and the primary star (\citealt{hol97})\vspace{-0.5em}
	\item effects of stellar encounter (\citealt{mal09}, whose simulations result in a range of separation $1\leq a\leq 6$ A.U.),\vspace{-0.5em}
\end{enumerate}
only migration can explain the short period planets and generally, the overall architecture of many systems.

The existence of the so-called \textit{hot Jupiters} (Jupiter-mass planets orbiting close to their host star, typically at  $a\leq 0.1 \rm{AU}$) augmented earlier speculations that such planets could be formed far away from the star and then, via an \textit{inward migration} process (\citealt{gold79,gold80,lin86II,lin86III,ward89}), which should be followed by an \textit{halting mechanism} (\citet{thommes03} implied that inclination excitation could serve as such and put, finally, an end in the migration), reached their present configurations.

Three mechanisms are believed to govern \textit{orbital evolution} \vspace{-0.5em}
\begin{itemize}
	\item \textit{gas disk migration} \vspace{-0.5em}
	\item \textit{planetesimal disk migration} \vspace{-0.5em}
	\item \textit{planet-planet scattering}. \vspace{-0.5em}
\end{itemize}
However, hot Earths, hot super Earths ($\sim 1-10 M_{\oplus}$) and hot Neptunes ($\sim 10-20 M_{\oplus}$) have been modeled by different scenarios (\citealt{ray08}).

Planet-planet scattering has, also, explained the occurrence of non-circular orbits with high orbital eccentricities (\citealt{mawei02,zt04,moad05}). Moreover, it can populate numerous orbital resonances (\citealt{raybag08}) and likewise, resonance chains ($4$:$2$:$1$ Laplace resonance) resulting from the damping mechanism of the \textit{planetesimal disk} (\citealt{ray10}). The damping force of the planetesimal disk can, furthermore, align the orbits in MMR through \textit{convergent migration}, if the planets of the system were formed in a stable configuration, namely, avoiding close encounters. 

Planetesimal disk migration (\citealt{malh95,tho08,arm07a}), in brief, causes an inward scattering of a planetesimal and an outward movement of the planet, or vice-versa, because of the conservation of the system's angular momentum. For a significant migration effect, the planetesimals' masses to be scattered should be of the order of the planet's mass. In the opposite case, planetesimal eccentricities are pumped up via dynamical friction and the planets remove them from the system, via dynamical ejection or collision. 

As for the solar system, migration seems to have affected its architecture and precisely, its outer regions (\citealt{gml04,tsi05,morlev08}). 

As far as the exosystems are concerned, large scale dynamical instabilities caused by such a planetesimal migration can interpret the highly eccentric orbits, often observed at them (see e.g. (\citealt{mawei02}), which defines, also, the so-called \textit{jumping Jupiters model}).

Gas disk migration (interactions between the planets and the residual gaseous protoplanetary disk) consists of three types: \vspace{-0.5em}
\begin{description}
   \item[I] It holds for planets of masses $M<10 M_{\oplus}$, whose evolution by wave excitation does not largely perturb the gas surface density profile, which in turn can be treated by using linear analysis (\citealt{war97}).\vspace{-0.5em}
   \item[II] It refers to larger planets. Their angular momentum dominates the disk's viscous forces, gas is repelled from their vicinity and an annular gap in the gas opens up at the orbital radius of the planets. A combined action of type II migration and planet-planet scattering has widely been proposed, e.g. by \citealt{moad05}.\vspace{-0.5em}
   \item[III or runaway migration] It is a more rapid migration, resulting from a modified, under certain circumstances (more massive disk), type I migration. The planets included in this type can open up a partial gap (see e.g. (\citealt{mapa03,pap05,paar06})). \vspace{-0.5em}
\end{description}

During the past two decades, it has broadly been approved that type II migration process can justify the resonant configurations of exoplanets (see e.g. (\citealt{haghi99,leepeal02,np02,kley03,pap03})). In two-planet systems, during resonance capture and given the planetary mass-ratio, the two planets follow particular migration \textit{paths}. These can be depicted on the eccentricities plane, when coplanar evolution is the case, either by the resonant stationary solutions of the averaged planetary three-body problem (see e.g. \citealt{mebeaumich03,lee04,bmfm06}), or by the resonant, stable, \textit{periodic orbits} of the general three-body problem (GTBP) in a suitable rotating frame. The families of periodic orbits constitute such paths, as shown in (\citealt{hadjvoy10,hv11}). A possible existence of such families in the general three-planet case is presented by \citet{v16}. 

\citet{thommes03} showed that stable spatial configurations can be obtained after migration and $2/1$ resonance capture, starting from a two-planet system of almost coplanar orbits. It was observed that when high values of the eccentricity were reached, excitation of the mutual inclination was taking place and hence, the system was additionally in ``inclination resonance''.  \citet{leetho09} showed numerically that when the outer planet was more eccentric than the inner the system jumped to nearby paths (from the symmetric to asymmetric families depending on the linear stability) and the inclination was excited. When along their evolution along the families (or paths) the two planets meet a VCO (\citealt{hen}) they abruptly get mutually inclined and they follow the generated spatial families of periodic orbits, as shown in \citet{vat14} for $2/1$ and $3/1$ MMRs. More precisely, therein, it was showcased that given the MMR and the planetary mass-ratio, we can tell whether the planets are likely to be mutually inclined or not.

In this study, we report the inclination excitation during the evolution due to type II migration process of circular orbits. We reveal the existence of -among others- $5/2$ and $7/3$ resonance captures imposed by the bifurcation points of the circular periodic orbits. In Sect. \ref{model}, we describe our model, the periodic orbits and their link with MMRs. In Sect. \ref{results}, we present our results and justify them through the computation of the periodic orbits of the circular family for greater values of multiplicity. 
Finally, we conclude in Sect. \ref{conc}.

\section{The model}\label{model}
We consider a system consisting of a star and two planets of masses $m_0$, $m_1$ and $m_2$, respectively, with $m_0\gg m_{1,2}$ where the indices 1 and 2 are always referring to the inner and the outer planet. Throughout the study the gravitational constant and the total mass of the system ($m=m_0+m_1+m_2$) are set equal to unity and we define the planetary mass-ratio as $\rho=\frac{m_2}{m_1}$. We investigate the dynamics in the framework of the GTBP, formulate the system in a suitable spatial rotating frame of reference as in (\citealt{av13}) and yield the Lagrangian of four degrees of freedom
\begin{equation}
\begin{array}{l}\resizebox{\textwidth}{!}{$
L=\frac{\displaystyle 1}{\displaystyle 2} \mu[a(\dot x_1^2+\dot z_1^2+x_1^2\dot \theta^2)+\displaystyle b [(\dot x_2^2+\dot y_2^2+\dot z_2^2)+\dot\theta^2(x_2^2+y_2^2)+2\dot\theta(x_2\dot y_2-\dot x_2y_2)]]-V,$}\nonumber
\label{Lagrangian}
\end{array}
\end{equation}
where $a=m_1/m_0$, $b=m_2/m$ and $V=-\frac{m_0 m_1}{r_{01}}-\frac{m_0 m_2}{r_{02}}-\frac{m_1 m_2}{r_{12}}$ is the potential with $r_{ij}$ indicating the distance between the bodies $i$ and $j$. 

Given the equations of motion, we can define the periodic orbits (periodic solutions) as the fixed or periodic points on a Poincar\'e surface of section through the repetitive crossings on that plane at $t=kT$, where $T$ is the period and $k$ the multiplicity of the orbit (see e.g. (\citealt{av12})). In cases where specific \textit{periodicity conditions} are fulfilled and two perpendicular crossings are established, the periodic orbits are called \textit{symmetric} and \textit{asymmetric} otherwise. 

The periodic orbits shape the phase space in their vicinity according to their linear stability (\citealt{hen,hadjbook06,sk01}) and can be either \textit{stable} or \textit{unstable}. Stable periodic orbits are surrounded by invariant tori and the long-term stability is guaranteed therein, whereas in the neighbourhood of the unstable ones chaotic domains exist which, in case the chaos is not weak, may destabilize the dynamical system. We should note that with regards to the planar periodic orbits the linear and vertical stability coexist but they do not affect each other. For instance, a planar unstable periodic orbit could be vertically stable if vertical deviations are considered and vice-versa, as seen in (\citealt{av12,av13,numan2014}). 

For a visual representation of the phase space and the delineation of the boundaries of the domains, maps of dynamical stability are utilized; for coplanar periodic orbits (\citealt{a16,av16,avHnl}) and mutually inclined orbits (\citealt{avv14,numan2014,a16}). It is straightforward that the families of stable periodic orbits constitute the backbone of stability domains.

In general, periodic orbits can be generated by particular types of bifurcation points (or critical orbits) and get mono-parametrically continued (see \citealt{hadj75,hen97,vkh09,avk11,av12,av13}). In this way, \textit{families} or \textit{characteristic curves} can be obtained, along which a specific parameter remains constant. However, there exist families that do not emanate from bifurcation points. For instance, depending on the planetary mass-ratio, there have been computed planar symmetric and asymmetric ones (\citealt{voyatzis08, vkh09, av16}), as well as, some symmetric mutually inclined orbits (\citealt{av12}).

The families can be classified as \textit{circular} or \textit{elliptic}. The former ones consist only of symmetric periodic orbits and along them the mean-motion ratio, $\frac{n_1}{n_2}$, varies, while the latter ones may consist of either symmetric or asymmetric ones, but the $\frac{n_1}{n_2}$ remains almost constant and is approximately equal to $\frac{p+q}{p}$, with $p,q\in\mathbb{Z}^*$ (c.c. the planetary case with the restricted one when the ratio is exactly equal to that rational number). Therefore, the elliptic families are resonant and the \textit{exact} MMR can be identified by the periodic orbits. 

In order for two-planets\footnote{Having assumed that $m_0\gg m_{1,2}$, the periodic orbits in the inertial frame correspond to almost Keplerian ellipses described by the \textit{osculating orbital elements} $a_i$ (semi-major axis), $e_i$ (eccentricity), $i_i$ (inclination), $\varpi_i$ (longitude of pericentre), $\Omega_i$ (longitude of ascending node) and $\lambda_i$ (mean longitude), where $i=1,2$.} to be locked in an MMR and evolve stably about periodic orbits, the resonant angles 
\begin{equation}\label{erangs}
\begin{array}{l}
\theta _{1}=(p+q)\lambda _{2}-p\lambda _{1}-q\varpi _{1}\\ 
\theta _{2}=(p+q)\lambda _{2}-p\lambda _{1}-q\varpi _{2}\\
\theta _{3}=(p+q)\lambda _{2}-p\lambda _{1}-\tfrac{q}{2}(\varpi _{1}+\varpi _{2})
\end{array}
\end{equation}
and the apsidal difference, $\Delta\varpi$, should librate about $0$ or $\pi$, if the orbit is symmetric, or around other angles, if it is asymmetric. The libration of the above resonant angles represents the \textit{eccentricity resonance} (e-resonance).

There exist four different symmetric configurations, if we assume aligned, $\Delta\varpi=0$, and anti-aligned, $\Delta\varpi=\pi$, planets, which do not change along the families of periodic orbits (see e.g. (\citealt{mbf06,av12,av13})). 

When the two planets are not coplanar, the resonant angles that define the \textit{inclination resonance} (i-resonance) for at least second order resonances  
\begin{equation}\label{irangs}
\begin{array}{l}
\varphi _{1}=(p+q)\lambda _{2}-p\lambda _{1}-q\Omega _{1}\\ 
\varphi _{2}=(p+q)\lambda _{2}-p\lambda _{1}-q\Omega _{2}\\
\end{array}
\end{equation}
librate as well. For instance, the inclination resonance of the 2/1 resonance would be the 4/2 (q = 2), since this is the lowest order of possible inclination resonance of this commensurability (see the arguments of the expansions of the disturbing function in \citet{murray}). 
We may further define the mixed resonance angle
\begin{equation}
\begin{array}{l}
\varphi _{12}=(p+q)\lambda _{2}-p\lambda _{1}-\Omega _{1}-\Omega_2=(\varphi_{11}+\varphi_{22})/q
\end{array}
\end{equation}
as well as the zeroth order secular resonance angle
\begin{equation}
\begin{array}{l}
\varphi _{\Omega}=\Omega _{1}-\Omega_2=(\varphi_{11}-\varphi_{22})/q.
\end{array}
\end{equation}

\section{Resonance capture and inclination excitation of circular orbits}\label{results}
In order to mimic the effects of type II migration process, we assume on the outer planet a Stokes type dissipative force (\citealt{bf93,bmfm06}) 
\begin{equation} \label{Fdissipative}
\mathbf{F}_d=-c(\mathbf{v}_p-\alpha\mathbf{v}_c)
\end{equation}
where $\mathbf{v}_p$ is the velocity of the planet and $\mathbf{v}_c$ is the circular velocity at the particular distance from the star. In a first order approximation, the constants $c$ and $\alpha$ are associated with the migration rate in semi-major axis, $\nu$, and the eccentricity damping, $K$, according to the formulae 
$$
\nu=2C(1-\alpha),\quad K=\frac{\alpha}{2(1-\alpha)}.
$$

In the numerical simulations we set the star mass $m_0=1 M_\odot$ and start with almost circular, $e_1=e_2=0.01$, and coplanar orbits, $i_1=i_2=0.1^\circ$, with $\Delta\Omega=180^\circ$. The inner planet is set to $a_1=5~$AU. 

\begin{figure}[h]
\centering
\includegraphics[width=7cm]{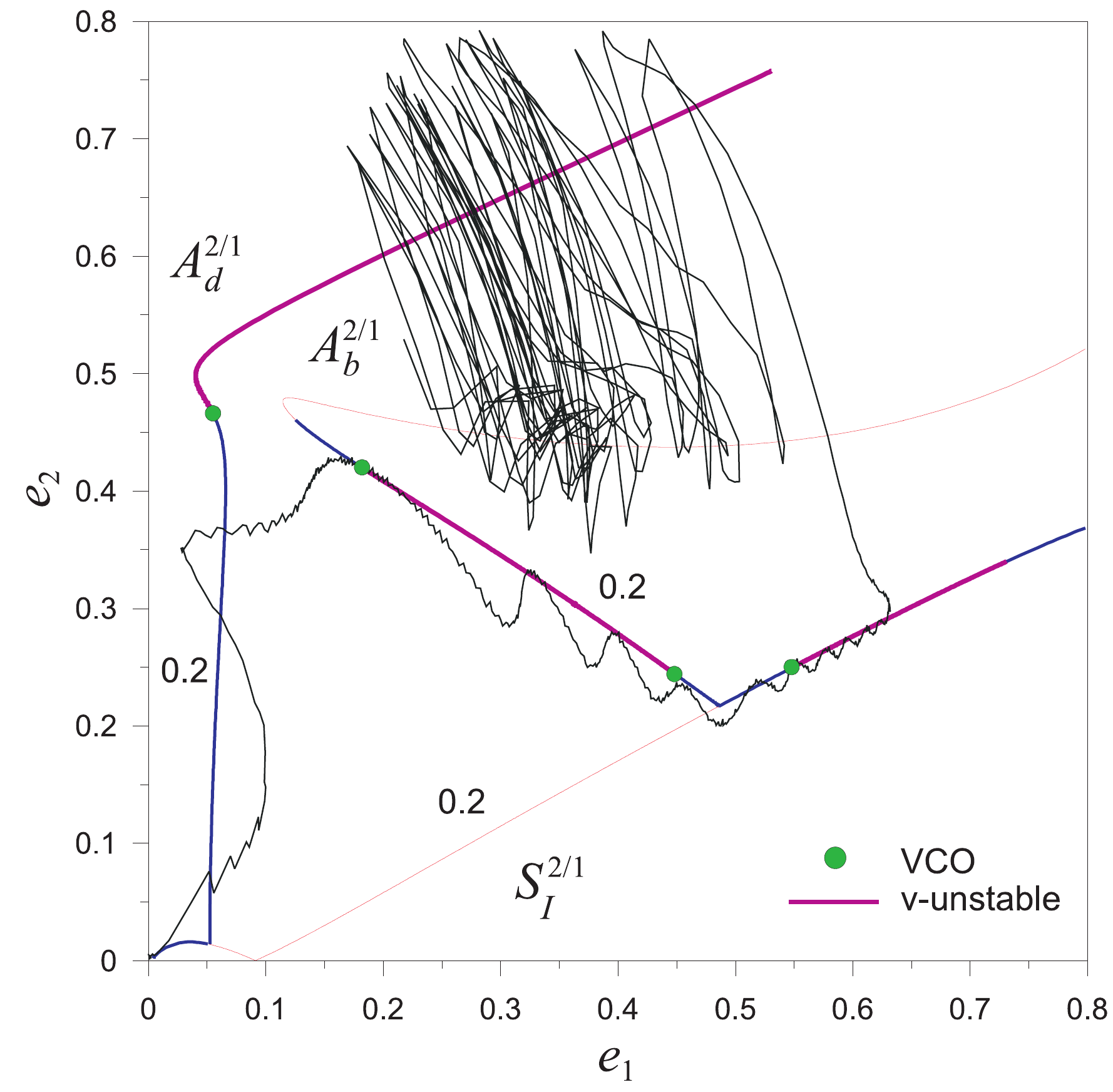}
\caption{Complicated phenomena for $\rho=0.2$ and $\nu=5.2~10^{-5}\textnormal{y}^{-1}$, $K=0.7$ in $2/1$ MMR. The evolution overcomes the asymmetric VCO of $A^{2/1}_b$ and then, the i-resonance appears, when the system reaches the next VCO located in the symmetric family,  $S^{2/1}_I$ .  Finally, the system jumps to another family of asymmetric periodic orbits,  $A^{2/1}_d$ . Blue bold lines refer to the stable periodic orbits, while the red coloured to the unstable ones. Magenta coloured lines correspond to vertical instability no matter the ``horizontal'' stability of the periodic orbits. The dots represent the VCOs.} 
\label{21n5k07r02_mig2d}
\end{figure}

\begin{figure}[h]
\centering
\includegraphics[width=\textwidth]{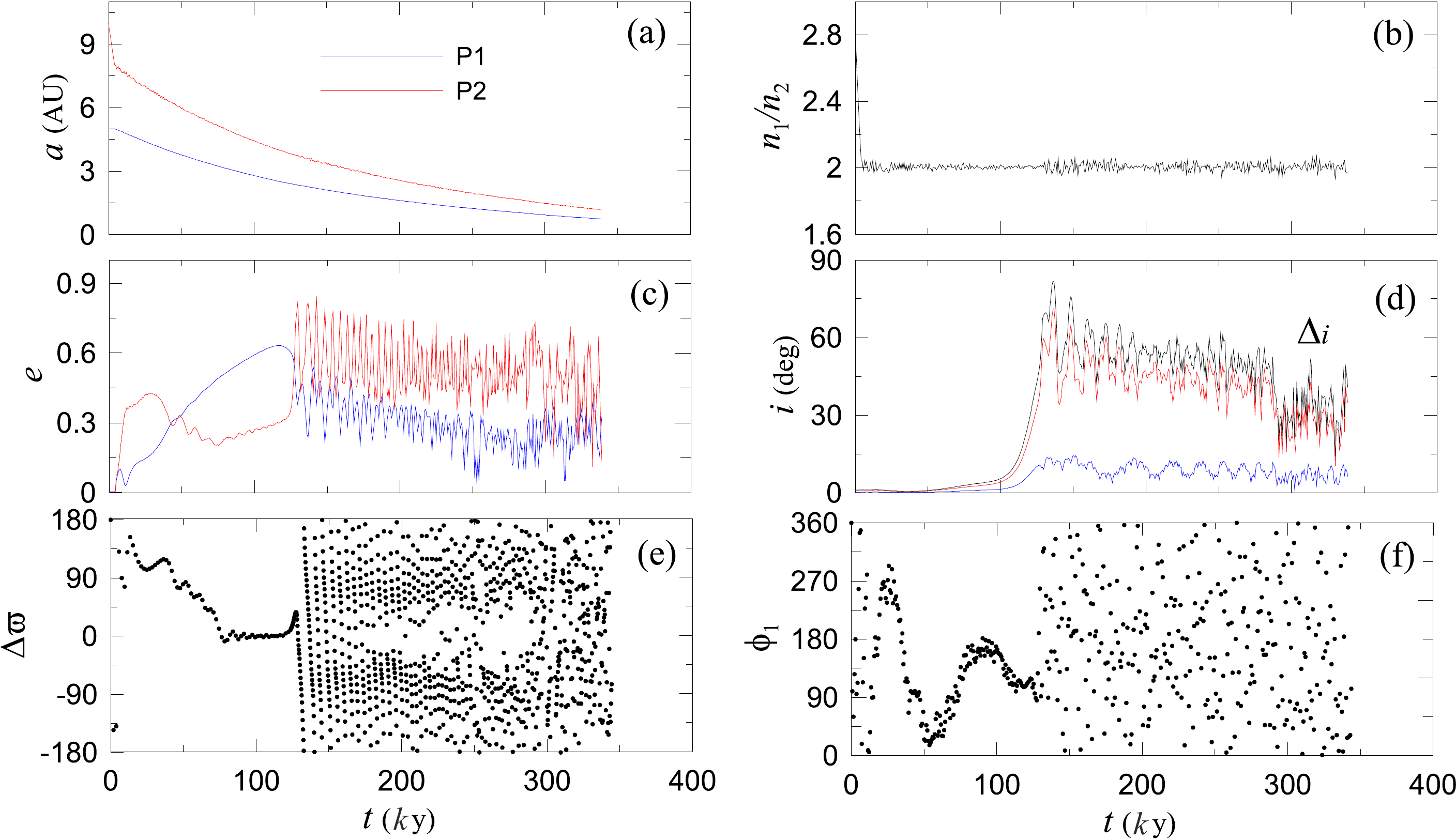}
\caption{Evolution of orbital elements under the influence of the dissipative force (Eq. \ref{Fdissipative}) with $\nu=5.2~10^{-5}\textnormal{y}^{-1}$, $K=0.7$ and planetary masses $m_1=1~M_J$, $m_2=0.2~M_J$.} 
\label{21n5k07r02_oe}
\end{figure}

As convergent migration proceeds, upon capture to e-resonance, the semi-major axes ratio will be $\frac{a_2}{a_1}\approx(\frac{n_1}{n_2})^{2/3}\approx(\frac{p+q}{p})^{2/3}$. At that moment, the planets will evolve along the elliptic\footnote{We note that, generally, the elliptic families bifurcate from the circular family at the orbits where $(\frac{n_1}{n_2})^{2/3}=(\frac{p+q}{p})^{2/3}$. Consequently, the planetary system evolves initially about the circular family up to the moment it gets captured in MMR.} family of stable periodic orbits that corresponds to their mass-ratio $\rho$. Then, both the resonant angles, Eq. \ref{erangs}, and the apsidal difference, $\Delta\varpi$, will librate about the angles that correspond to the particular configuration of the periodic orbits they follow. Along their evolution, when a VCO is met, the planets follow the spatial family that bifurcates from it. Then, i-resonance is observed in addition to e-resonance, via the libration of the resonant angles of Eq. \ref{irangs}. This scenario is showcased by \citet{vat14}. Therein, given the MMR and the $\rho$, the inclination excitation took place at either asymmetric or symmetric VCO and this type of evolution has been verified for $\nu\lesssim 10^{-6}\,y^{-1}$ and for small eccentricity damping ($K\approx 1$). We have also observed complicated phenomena discussed in \citet{leetho09} (see e.g. Figs. \ref{21n5k07r02_mig2d} and \ref{21n5k07r02_oe}).

\subsection{5/2 resonance capture}

\begin{figure}[h]\centering
\includegraphics[width=10cm]{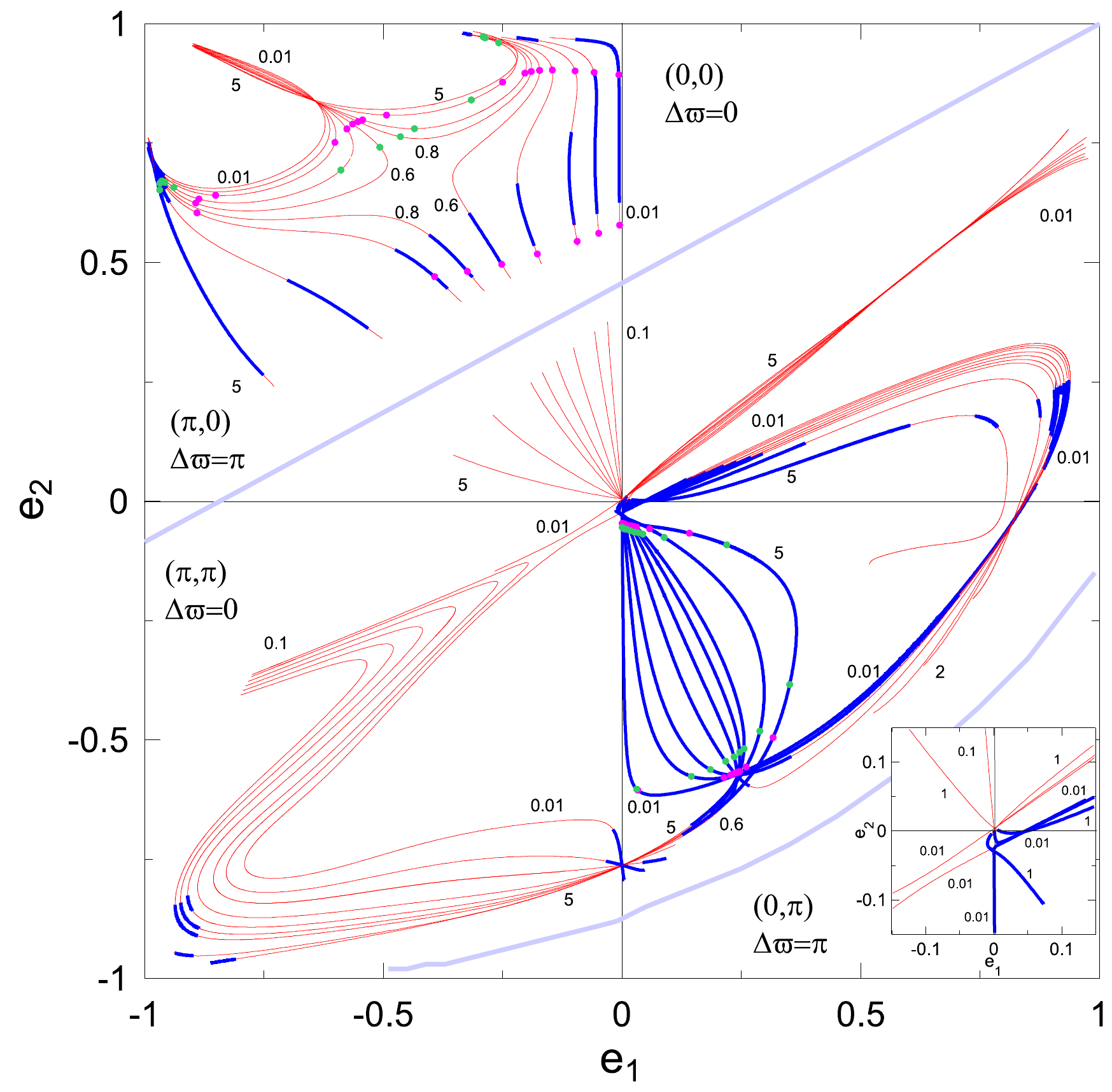}  
\caption{The planar families of $5/2$ resonant periodic orbits. Presentation as in Fig. \ref{21n5k07r02_mig2d}. The bold gray lines correspond to regions of close encounters and collisions and the coloured dots depict the VCOs. The numbers along the families represent the planetary mass-ratio $\rho$, where $m_1=1~M_J$.} 
\label{52again}
\end{figure}

In Fig. \ref{52again},  we present the planar families of symmetric periodic orbits in the $(e_1,e_2)$ plane. The four panels correspond to the four different configurations as attributed by the resonant angles $(\theta_1, \theta_2)$ and the apsidal difference, $\Delta\varpi$. The negative values of the eccentricities are used whenever $\theta_{1,2}$ librates about $\pi$. For more details see (\citealt{av13}).

\begin{figure}[h]
\centering
\includegraphics[width=\textwidth,height=7.5cm]{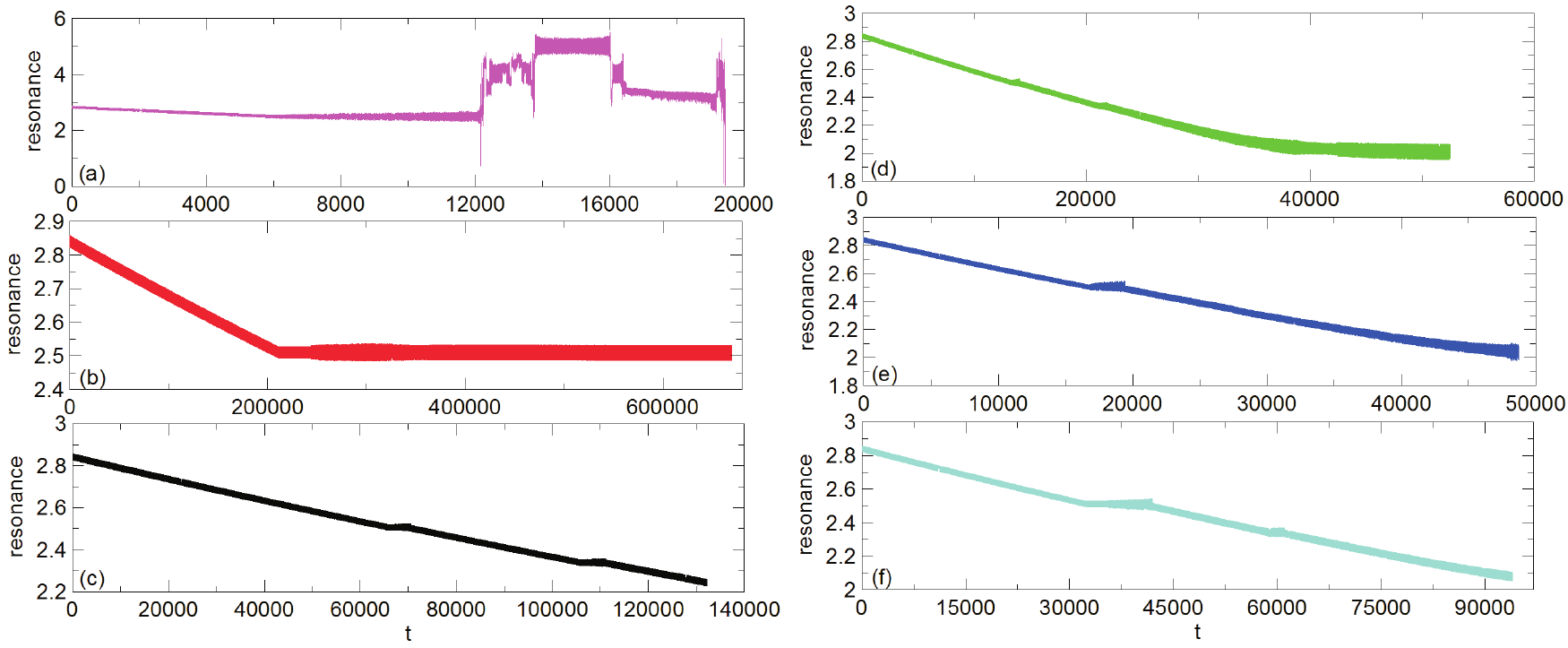}
  \caption{Captures in $5/2$ MMR. {\bf a} A case where the eccentricity damping is $K=4.5$, the migration rate is  $\nu=1.26$ $10^{-8}$ $y^{-1}$ and $\rho=20$. {\bf b} The critical case of $\alpha =1$, namely the migration rate in semi-major axis equals zero. The mass ratio of the planets is $\rho=5$. {\bf c} A case where $K=4.5$, $\nu=1.26$ $10^{-9}$ $y^{-1}$ and $\rho=2$. {\bf d} A case where $K=0.5$, $\nu=6.28$ $10^{-9}$ $y^{-1}$ and $\rho=5$. {\bf e}  A case where $K=0.75$, $\nu=5.03$ $10^{-9}$ $y^{-1}$ and $\rho=5$. {\bf f}  A case where $K=2$, $\nu=2.5$ $10^{-9}$ $y^{-1}$ and $\rho=5$.}
  \label{res}
\end{figure}

\begin{figure}[h]
\centering
\begin{tabular}{c@{\hspace*{-6pt}}c}
\begin{minipage}[t]{.5\textwidth}
  \hspace*{-10pt}\includegraphics[width=6cm,height=10cm]{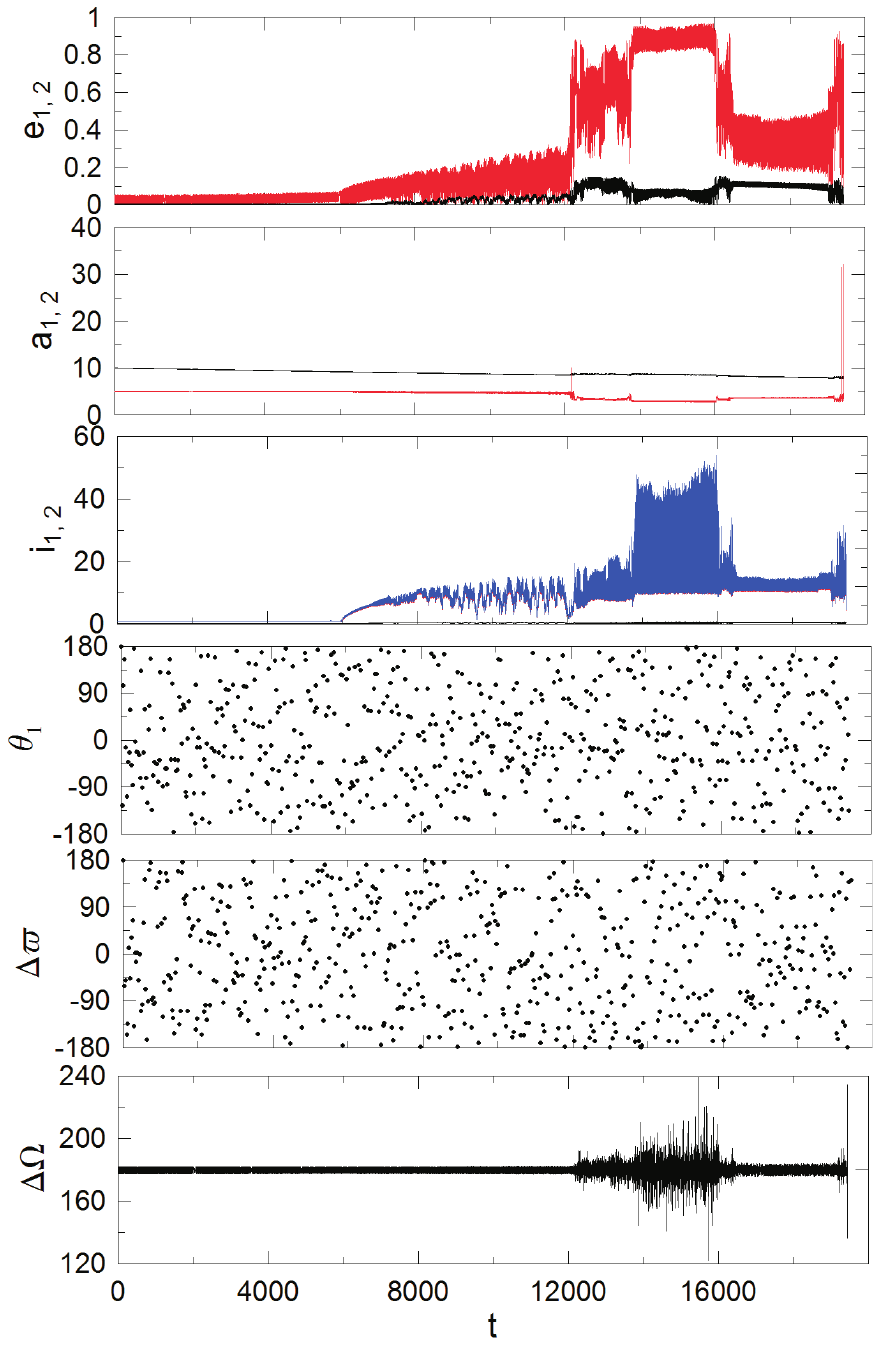}
  \caption{Evolution of orbital elements and resonant angles of a planetary system described in Fig. \ref{res}a. Red coloured lines stand for the inner planet, black ones for the outer planet and the blue one for mutual inclination.}
  \label{42}
\end{minipage}%
&
\begin{minipage}[t]{.5\textwidth}
  \includegraphics[width=6cm,height=10cm]{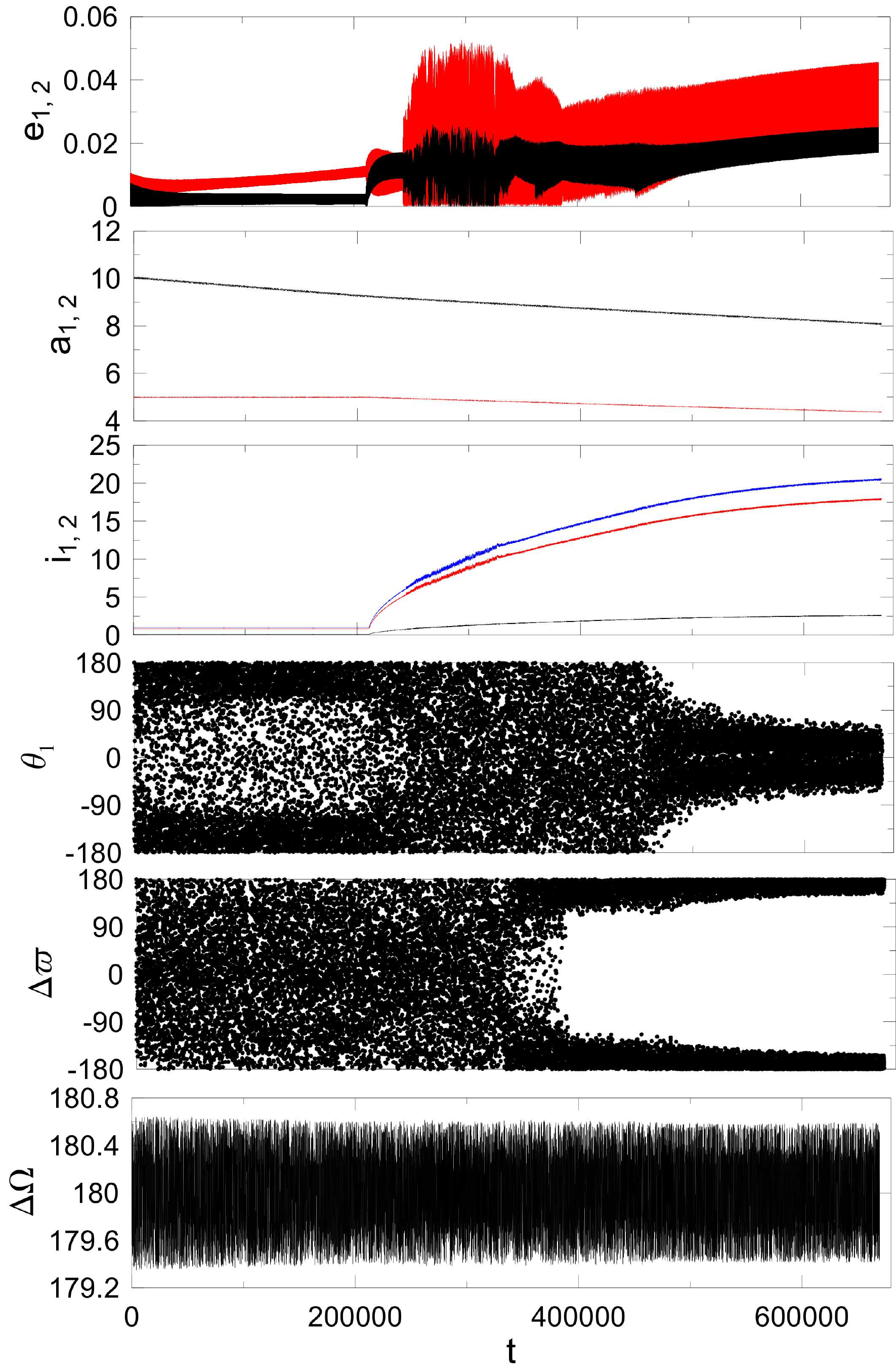}
  \caption{Evolution of a planetary system described in Fig. \ref{res}b presented as in Fig. \ref{42}.}
  \label{37}
\end{minipage}
\end{tabular}
\end{figure}

\begin{figure}[h]
\centering
\begin{tabular}{c@{\hspace*{-6pt}}c}
\begin{minipage}[t]{.5\textwidth}
  \hspace*{-10pt}\includegraphics[width=6cm,height=10cm]{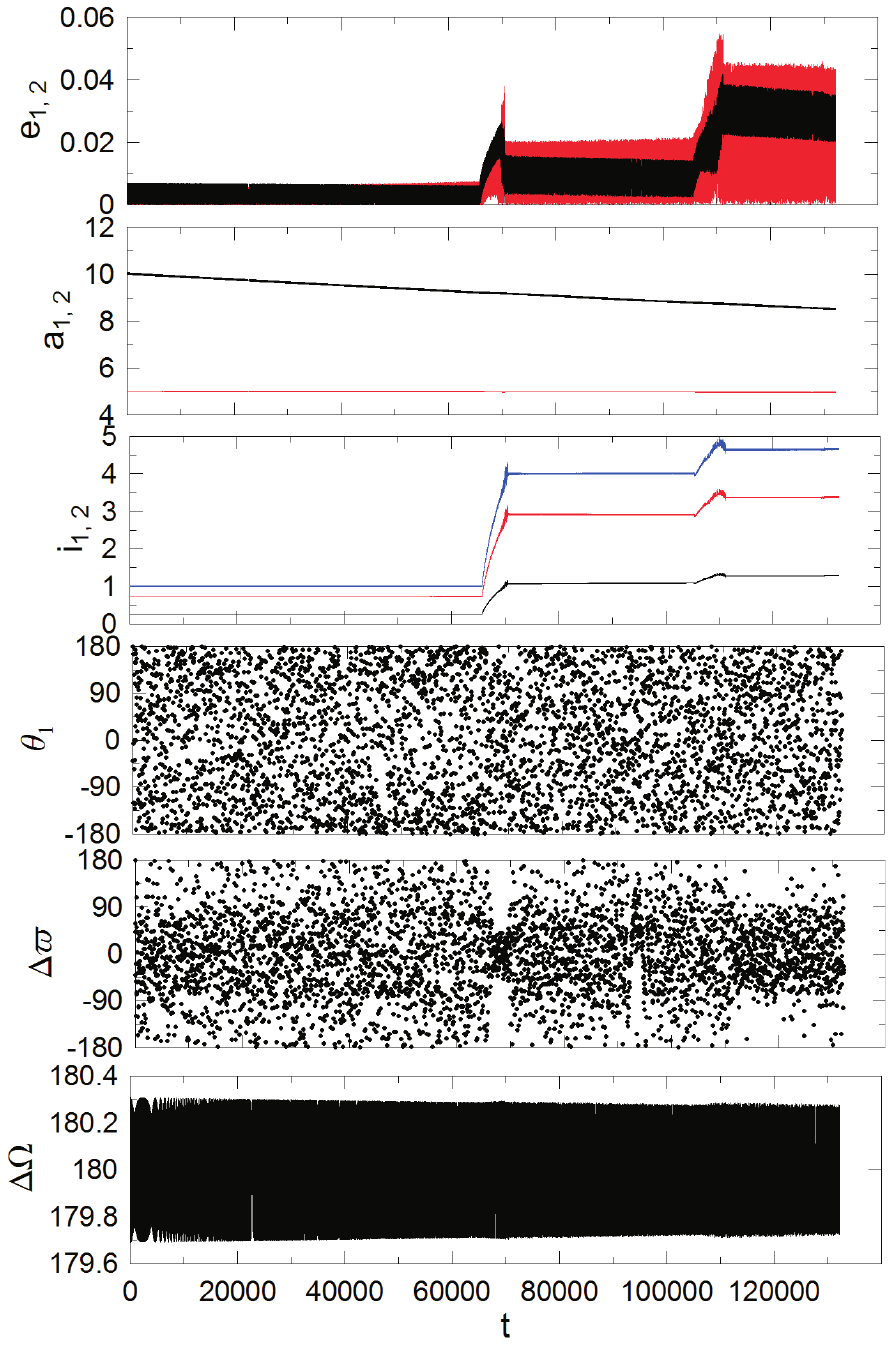}
  \caption{Evolution of a planetary system described in Fig. \ref{res}c presented as in Fig. \ref{42}.}
  \label{3black}
\end{minipage}%
&
\begin{minipage}[t]{.5\textwidth}
  \includegraphics[width=6cm,height=10cm]{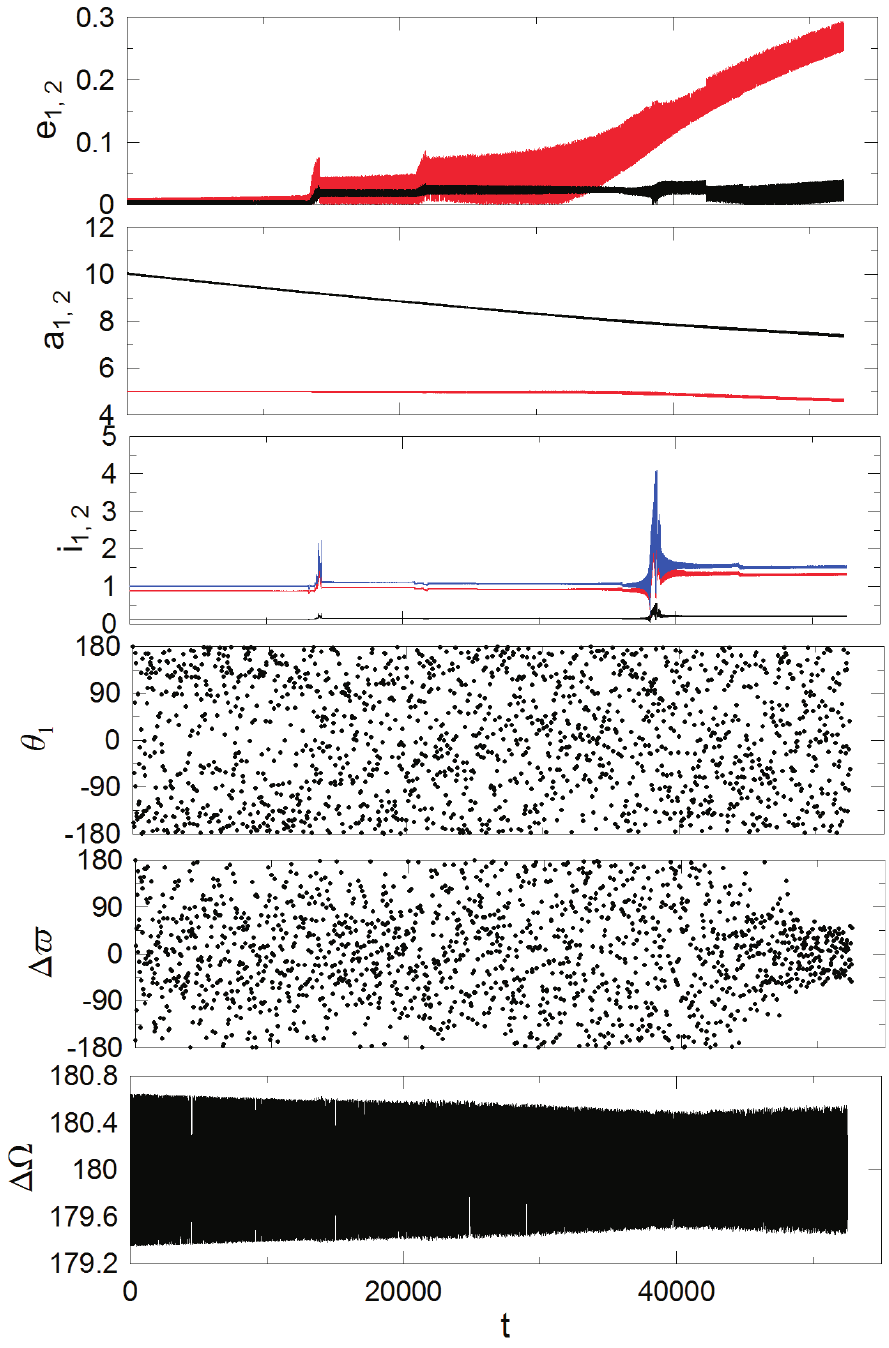}
  \caption{Evolution of a planetary system described in Fig. \ref{res}d presented as in Fig. \ref{42}.}
  \label{4green}
\end{minipage}
\end{tabular}
\end{figure}

\begin{figure}[h]
\centering
\begin{tabular}{c@{\hspace*{-6pt}}c}
\begin{minipage}[t]{.5\textwidth}
  \hspace*{-10pt}\includegraphics[width=6cm,height=10cm]{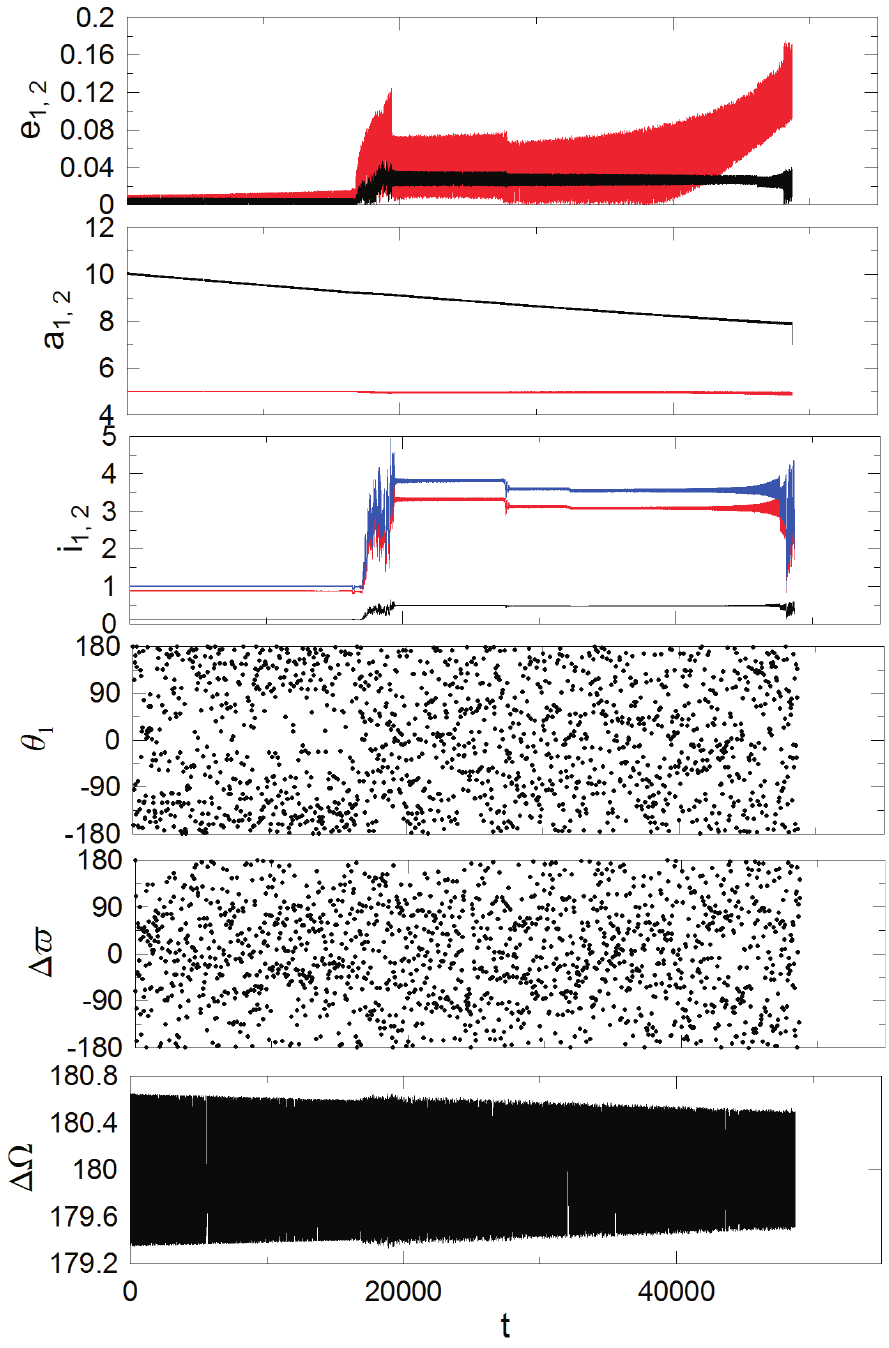}
  \caption{Evolution of a planetary system described in Fig. \ref{res}e presented as in Fig. \ref{42}.}
  \label{12blue}
\end{minipage}%
&
\begin{minipage}[t]{.5\textwidth}
  \includegraphics[width=6cm,height=10cm]{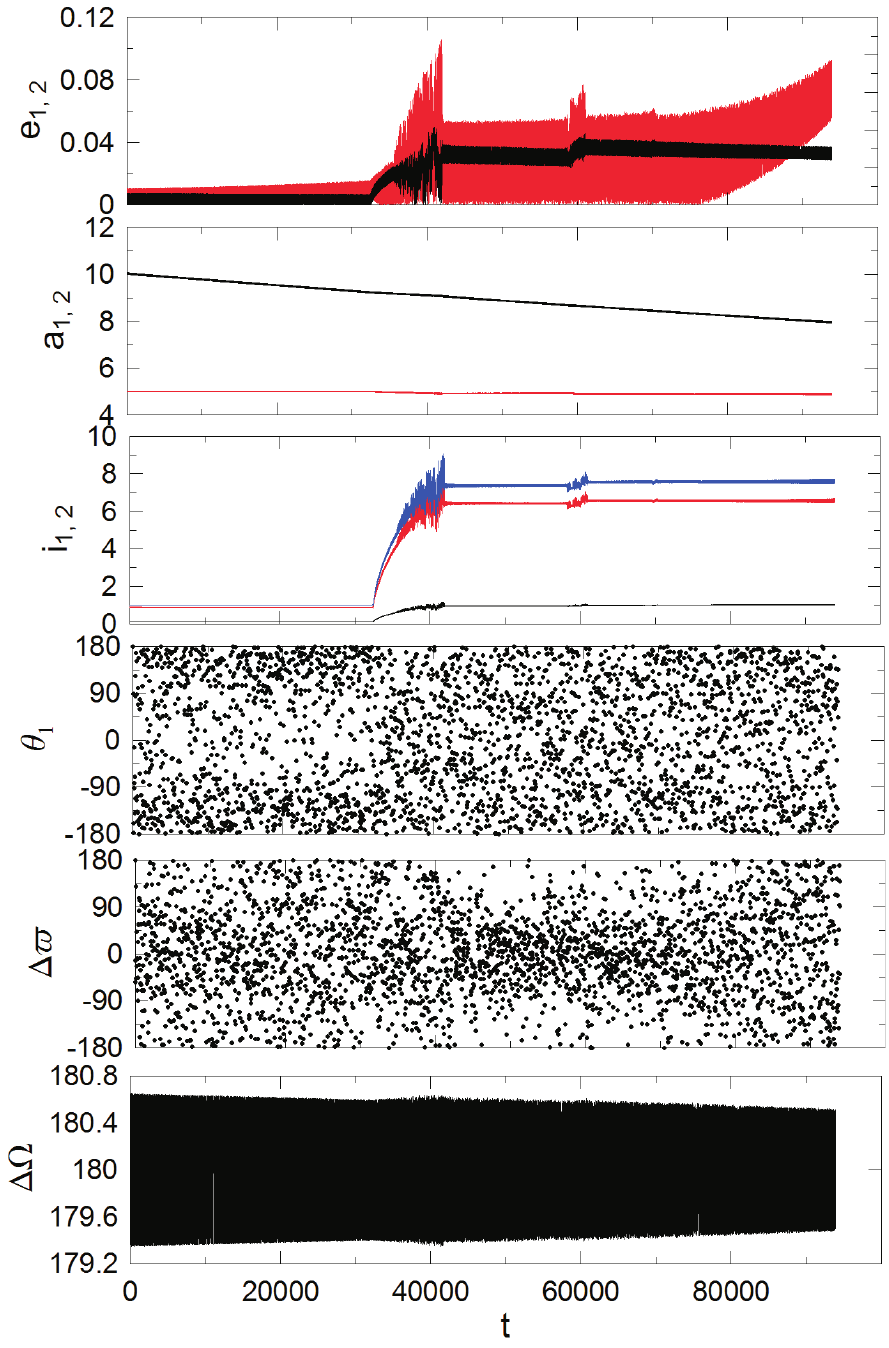}
  \caption{Evolution of a planetary system described in Fig. \ref{res}f presented as in Fig. \ref{42}.}
  \label{11cyan}
\end{minipage}
\end{tabular}
\end{figure}

In the quest to achieve $5/2$ resonance capture, as reported by \citet{litsi09}, we examined all possible combinations of values $10^{-9}\leq \nu \leq 10^{-5}$ ($y^{-1}$) and $K\leq 1$ for mass ratios $0.1\leq \rho \leq 5$. All of the systems of planetary masses with $\rho<2$ passed by $5/2$ MMR without being captured and ended up locked in $2/1$ MMR. The only cases of systems that have been captured in $5/2$ MMR along with the parameters used are shown in Fig. \ref{res}. Then, the majority of them (apart from the critical case in Fig. \ref{res}b) was temporarily trapped in $7/3$ and finally in $2/1$ MMR. In Figs. \ref{42}-\ref{11cyan}, we present the evolution of the eccentricities, semi-major axes, inclinations and resonant angles, $\theta_1$, $\Delta \varpi$ and $\Delta \Omega$, of the planets. 

The path that the planar families of $5/2$ MMR follow, when the eccentricities are approximately equal to zero  (see Fig. \ref{52again}), could also aid to justifying the non-capture  to e-resonance, even for lower planetary mass-ratios. Nevertheless, we observe that even when the mutual inclination of the planets increased significantly, the planetary orbits remained circular. The above mentioned behaviour -temporal captures in $5/2$ and $7/3$ MMRs and a final capture in $5/2$ MMR only for a specific choice of a parameter- is fully explained in Sect. \ref{just52}. 

The planets of the system forced to evolve under the parameters given in Fig.~\ref{res}a were firstly trapped in $5/2$\;MMR, then in $4/1$,\;$5/1$ again in\;$4/1$ and finally, in $7/2$ MMR. The planets reached the maximum mutual inclination and eccentricity values during $5/1$ capture, but in the end, the system was distorted, since the initially inner planet escaped (see also Fig. \ref{42}).

The planets of the system described in Fig. \ref{res}b evolved under a critical case of no migration rate in semi-major axis and obtained high mutual inclination values of a symmetric stable configuration of anti-aligned planets (libration of $\theta_1$ about $0^{\circ}$, $\Delta \varpi$ about $\pi$ and $\Delta \varpi$ about $\pi$). However, their orbits were circular ($e_i<0.06$) throughout the simulation (see Fig. \ref{37}).

The evolution of the systems described in Figs. \ref{res}c,f is quite similar, i.e. the planets, after having been captured in $5/2$ MMR, reached even higher mutual inclination values, when they were locked in $7/3$ MMR. Both their eccentricities increased after these consecutive captures, although none of them surpassed the value $0.1$, namely they remained on circular orbits (see Figs. \ref{3black} and \ref{11cyan}, respectively). 

The planets of the system described in Fig. \ref{res}d remained almost planar throughout the evolution, even when they were trapped in $2/1$ MMR and the eccentricity of the inner planet started increasing significantly (see also Fig. \ref{4green}). They, also, ended up in a symmetric configuration ($\Delta \varpi$ librated about $0^{\circ}$), however, the motion was unstable ($\theta_1$ rotated). 

The evolution of the system of parameters shown in Fig. \ref{res}e is similar to the previous one with regards to the evolution of the orbital elements and the resonant angles. They solely differ in the relative longitude of periapse, $\Delta \varpi$, which did not finally librate, but kept on rotating instead (see Fig.~\ref{12blue}).

\begin{figure}[h]
\includegraphics[width=\textwidth]{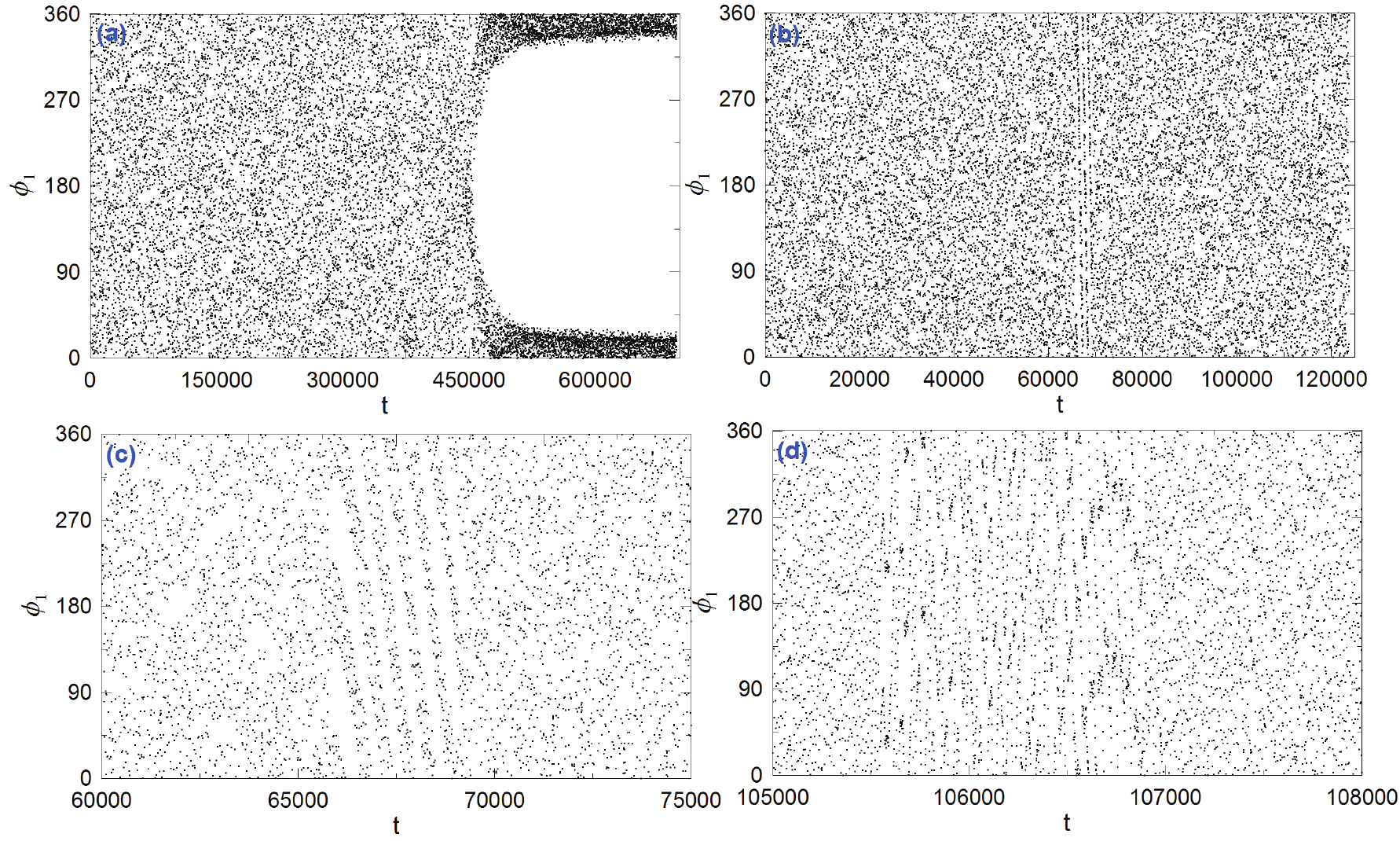}  
\caption{The evolution of the resonant angle $\varphi_1$, for the systems of Figs. \textbf{a} \ref{37} and \textbf{b} \ref{3black}. Magnifications of \textbf{b} are provided in \textbf{c} and \textbf{d}.} 
\label{phi152}
\end{figure}

In Fig. \ref{phi152}a, we present the inclination resonant angle $\varphi_1$ and observe what was expected for the system in Fig. \ref{37}; when the i-resonance was achieved, it started librating around $0^\circ$, as the spatial family the system followed consists of $xz$-symmetric periodic orbits (see (\citealt{av12}) for the definition of this type of orbits). In Fig. \ref{phi152}b, we show the evolution of $\varphi_1$ in the system described in Fig. \ref{3black}, where firstly $5/2$ MMR capture is apparent and then, a $7/3$ temporal capture is seen. In Figs. \ref{phi152}c and d, we show magnifications of the regions, where those limited in time captures were encountered.

\subsubsection{Justification of 5/2 resonance capture}\label{just52}

The $5/2$ resonance capture could not be explained with the aid of the VCOs of the elliptic families of periodic orbits of the GTBP, because they arise for larger values of eccentricities (see the magnification in Fig. \ref{52again} herein and Fig. 5 in \citet{av13}). 

Thus, we sought for their existence in the circular family. When the mulptiplicity of those periodic orbits is equal to 1, the only VCOs that exist, belong to $3/1$ MMR (see Fig. 3 of (\citealt{vat14})). Consequently, we had to compute the circular family of periodic orbits for larger values of multiplicity. 

\begin{figure}[h]
$$
\begin{array}{c}
\includegraphics[width=0.75\textwidth]{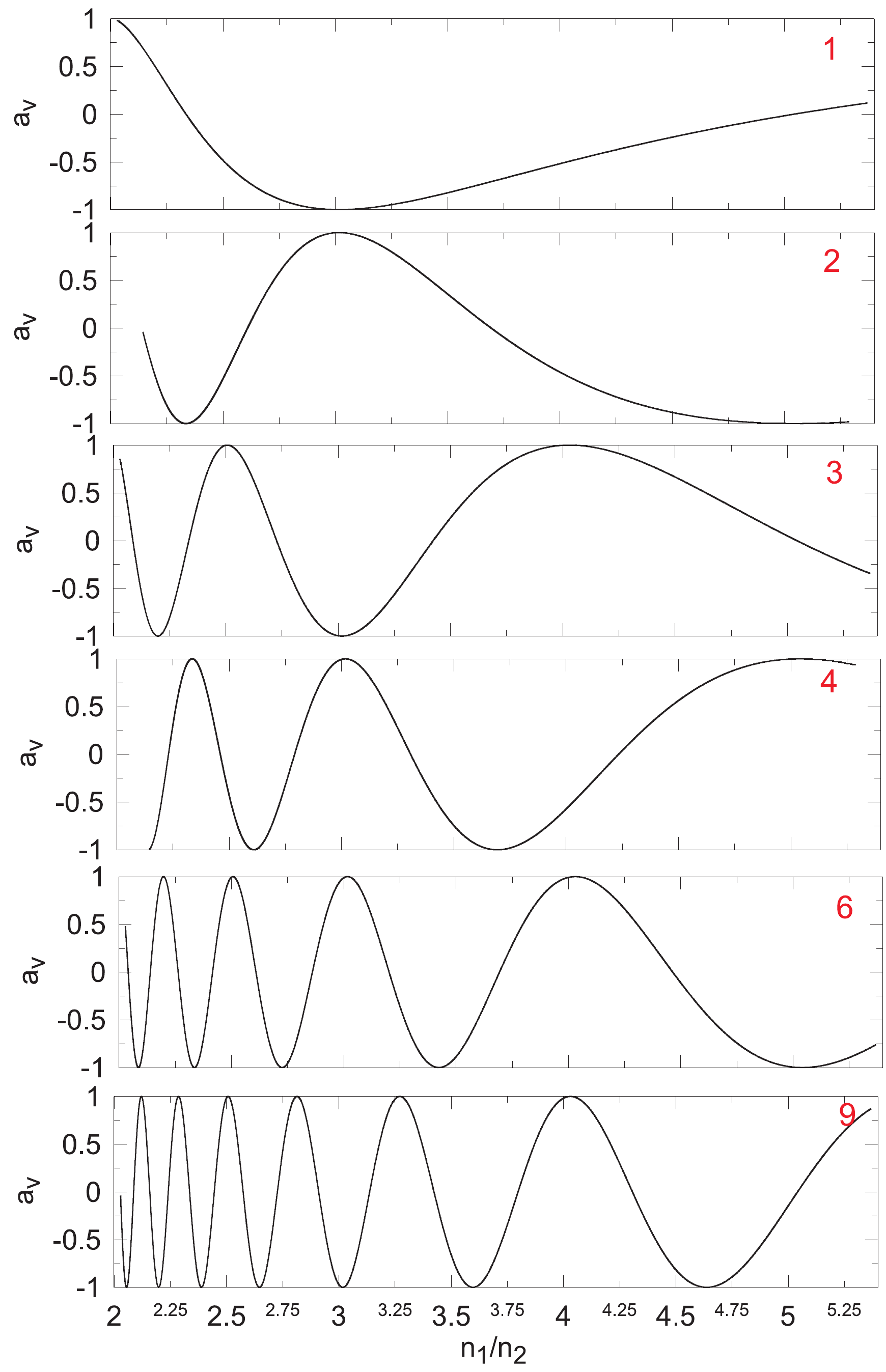}
\end{array}$$
\caption{The vertical stability index, $a_v$, along the circular family as the multiplicity (bold red numbers at top right corner of each panel) of the periodic orbits increases and the mean-motion ratio, $n_1/n_2$, varies.} 
\label{cfamdifmul}
\end{figure}

In Fig. \ref{cfamdifmul}, we present the variation of the vertical stability index as the multiplicity of periodic orbits increases for MMRs greater than $2/1$. The VCOs which appear along the family at particular MMRs with respect to the multiplicity values are shown in Table \ref{multab}.

\begin{figure}[h]
$$
\begin{array}{c}
\includegraphics[width=8cm]{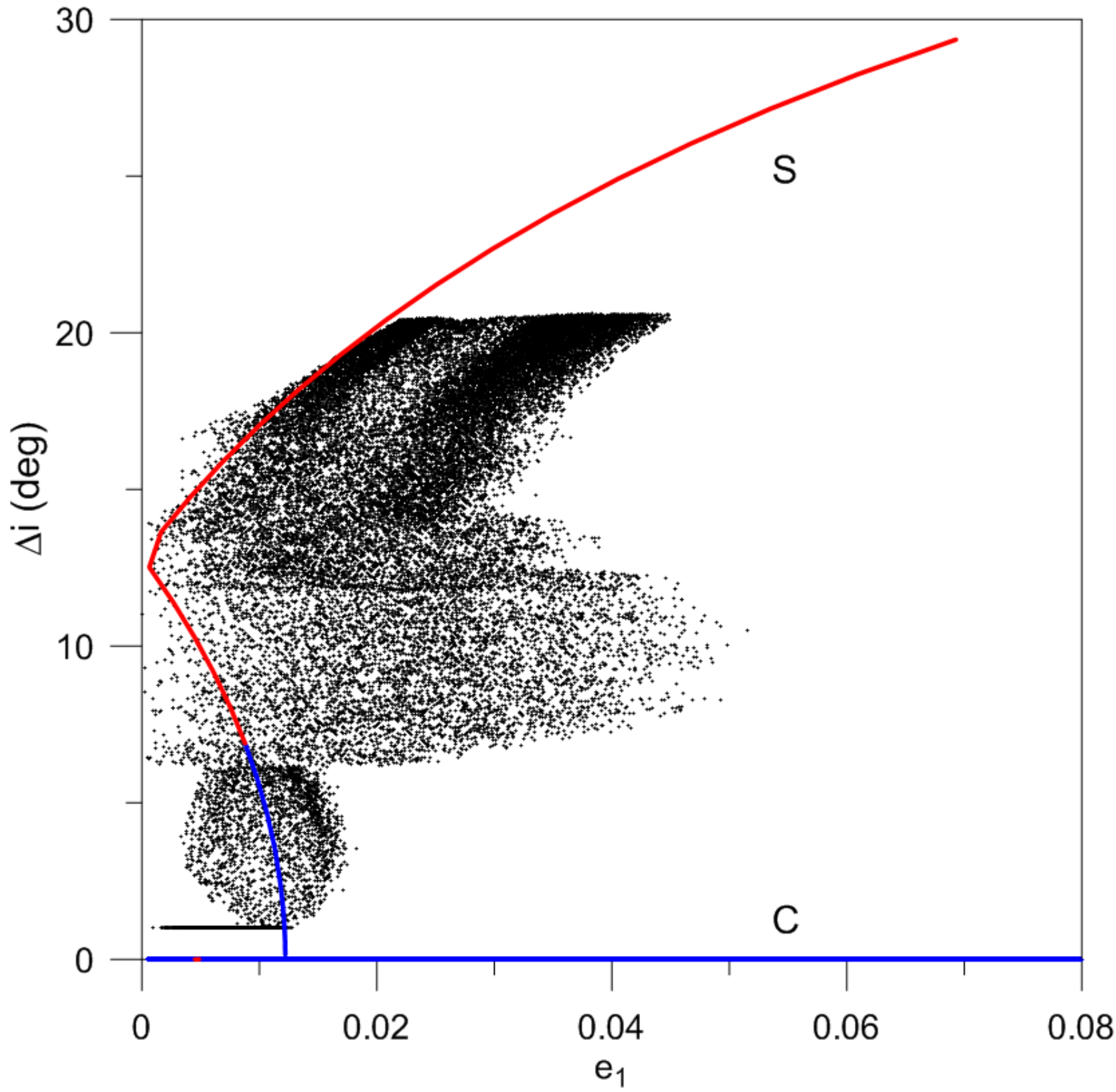}
\end{array}$$
\caption{Capture in $5/2$ MMR and evolution (black dots) along a spatial symmetric elliptic family, $S$, which bifurcated from the circular, $C$, one. The mutual inclination, $\Delta i$,  increased abruptly, when the VCO was met. Blue bold lines refer to the stable periodic orbits, while the red coloured to the unstable ones.} 
\label{523dcircmig}
\end{figure}

In Fig. \ref{523dcircmig}, we project the evolution of the migrating system described\;in Fig. \ref{37} in the plane of $(e_1,\Delta i)$ for visualization reasons. We observe the\;ex\-citation of the orbit from its circular planar form to planar elliptic (e-resonance) and to inclined (i-resonance) motion. The system followed the family of spatial periodic orbits, which was generated by the VCO of the\;cir\-cular family derived by a periodic orbit of multiplicity equal to 3 for $\rho=5$. The larger oscillations start taking place as the system enters the unstable region in the vicinity of the unstable periodic orbits in phase space.

\begin{sidewaystable}[h]
\begin{center}
\caption{We provide the MMRs that correspond to the appearance of VCO along the circular family as the multiplicity of periodic orbits, with which they are computed, increases.}
\resizebox{\textwidth}{!}{\begin{tabular}[b]{ccccccccccccccccccccc}
\hline\hline
Multiplicity & \multicolumn{19}{c}{ Mean-motion resonance}\\
\hline
1&&&&&&&&&&&&3/1&&&&&&&\\
2&&&&&7/3&&&&&&&3/1&&&&&&&5/1\\
3&&&11/5&&&&5/2&&&&&3/1&&&&&4/1&&\\
4&&&&&7/3&&&13/5&&&&3/1&&&&11/3&&&5/1\\
6&&17/8&11/5&&7/3&&5/2&&&19/7&&3/1&&17/5&&&4/1&&5/1\\
9&2/1&17/8&11/5&16/7&&12/5&5/2&&8/3&&14/5&3/1&10/3&&18/5&&4/1&23/5&\\
\hline
\end{tabular}}\label{multab}
	\end{center}
\end{sidewaystable}

\section{Conclusions}\label{conc}

In this study, we performed a direct link of the inclination excitation of two-planet systems on circular orbits with the intrinsic property of the three-body problem: the vertical critical orbits. 

It was shown that during type II migration process and after capture in MMR, the VCOs of the circular family can provide the means to i-resonance. The generated spatial families of periodic orbits of high multiplicity constitute the paths that guide the evolution of planetary systems. 

The circular family can pave the way to the appearance of more VCOs in different MMRs, as the periodic orbits, it consists of, are computed with an increased multiplicity. Then, possible trappings in the corresponding MMRs can be justified and the planets can get mutually inclined by following the respective stable, spatial elliptic families.

\acknowledgements{The work of KIA was partially supported by the Fonds de la Recherche Scientifique-FNRS under Grant No. T.0029.13 (``ExtraOrDynHa'' research project).}

\clearpage

\bibliographystyle{aasjournal}

\bibliography{nbib}

\end{document}